\documentclass[prl,aps,twocolumn,superscriptaddress]{revtex4}
\usepackage{latexsym}
\usepackage{graphicx}

\newcommand{\be}{\begin{equation}}
\newcommand{\ee}{\end{equation}}
\newcommand{\ba}{\begin{eqnarray}}
\newcommand{\ea}{\end{eqnarray}}
\newcommand{\baa}{\begin{eqnarray*}}
\newcommand{\eaa}{\end{eqnarray*}}

\def\be{\begin{equation}}
\def\ee{\end{equation}}
\def\bea{\begin{eqnarray}}
\def\eea{\end{eqnarray}}

\def\C60{A$_x$C$_{60}$}

\def\HgCu3{HgCa$_2$Cu$_3$O$_{8+y}$}
\def\HgCu4{HgBa$_2$Ca$_3$Cu$_4$O$_{10+y}$}
\def\TlCu{Tl$_2$Ba$_2$CuO$_{6+\delta}$}
\def\TlCu3{Tl$_2$Ba$_2$Ca$_2$Cu$_3$O$_{10+y}$}
\def\TlCu4{Tl$_2$Ba$_2$Ca$_3$Cu$_4$O$_{12+y}$}

\def\BiCu3{Bi$_2$Sr$_2$Ca$_{2}$Cu$_3$O$_y$}

\def\8LSCO{La$_{1.88}$Sr$_{.12}$CuO$_4$}
\def\110LNSCO{La$_{1.5}$Nd$_{0.4}$Sr$_{0.1}$CuO$_{4}$}
\def\stage4LCO{La$_{2}$CuO$_{4+\delta}$}
\def\Y248{YBa$_2$Cu$_4$O$_8$}

\def\NbSe2{NbSe$_2$}
\def\TaSe2{TaSe$_2$}
\def\TiSe2{TiSe$_2$}

\begin{document}

\title{ Magnetic Frustration and Iron-Vacancy Ordering in Iron-Chalcogenide }
\author{Chen Fang}
\affiliation{Department of Physics, Purdue University, West
Lafayette, Indiana 47907, USA}
\author{Bao Xu}
\affiliation{Beijing National Laboratory for Condensed Matter
Physics, Institute of Physics, Chinese Academy of Sciences, Beijing
100080, China}
\author{Pengcheng Dai}
\affiliation{The University of Tennessee, Knoxville, Tennessee 37996-1200, USA}
\affiliation{Beijing National Laboratory for Condensed Matter
Physics, Institute of Physics, Chinese Academy of Sciences, Beijing
100080, China}
\author{Tao Xiang}
\affiliation{Beijing National Laboratory for Condensed Matter
Physics, Institute of Physics, Chinese Academy of Sciences, Beijing
100080, China}
\affiliation{  Institute of Theoretical Physics, Chinese Academy of Sciences, Beijing
100080, China}
\author{Jiangping Hu}
\affiliation{Department of Physics,
Purdue University, West Lafayette, Indiana 47907, USA}
\affiliation{Beijing National Laboratory for Condensed Matter
Physics, Institute of Physics, Chinese Academy of Sciences, Beijing
100080, China}

\begin{abstract}
We show that  the magnetic and vacancy  orders in the  122 $(A_{1-y}Fe_{2-x}Se_2)$ iron-chalcogenides can be  naturally derived from the $J_1-J_2-J_3$ model  with  $J_1$  being the ferromagnetic (FM) nearest neighbor exchange coupling and $J_{2}, J_3$ being  the antiferromagnetic  (AFM) next and third nearest neighbor ones respectively, previously proposed to describe the magnetism in the 11(FeTe/Se) systems.  In the 11 systems,   the magnetic exchange couplings are extremely frustrated  in the ordered  bi-collinear antiferromagnetic state so that the  magnetic transition temperature is low.   In the 122 systems, the formation of iron vacancy order  reduces the magnetic frustration and significantly increases the magnetic transition temperature and the ordered magnetic moment.  The pattern of the 245 iron-vacancy order ($\sqrt{5}\times \sqrt{5}$) observed in experiments is correlated to  the maximum reduction of  magnetic frustration.  The nature of the iron-vacancy  ordering  may hence be electronically driven.  We explore other possible  vacancy patterns and magnetic orders associated with them.  We also calculate the spin wave excitations and their novel features to test our model.
\end{abstract}

\maketitle

The recent discovery of a new family of iron-based superconductors,
the 122 iron-chalcogenides
$A(K,Cs,Rb)_yFe_{2-x}Se_2$\cite{Guo2010,Fang2010,Liu2011},  with a
superconducting transition temperature  even higher than 40 K  has
attracted many research attentions. The compounds are heavily
electron doped with only electron Fermi pockets which are mainly
located at the $M$ point of the unfolded Brillouin zone  as shown by
both angle-resolved photoemission spectroscopy
(ARPES)\cite{ZhangY2010,WangXP2011,Mou2011} and LDA
calculations\cite{Cao2010,ZhangL2009,Yan2010}.   Among all of
iron-based superconductors, the new materials  also have the highest
magnetic transition temperature  around 500K and the largest ordered
magnetic moment  around $3\mu_b$\cite{Bao2011a,Bao2011b}.   The
materials are also featured with intrinsic iron vacancies which can order themselves below some transition temperatures\cite{Bao2011a,Bacsa2011,Pomjakushin2011b,WangZ2011,Zavali2011,ZhangAM2011}.
Currently, it is unclear how the magnetic, vacancy and
superconducting orders are correlated.

For iron-pnicitides, their parental compounds display an universal
collinear antiferromagnetic (CAFM) phase\cite{Cruz2008},  which can
be described by a simple magnetic exchange model including the
nearest neighbor (NN) $J_1$ (note,  in magnetically ordered state,
the effective exchange coupling becomes $J_{1a}$ and $J_{1b}$ due to
rotation symmetry breaking)
 and the next nearest neighbor(NNN) $J_2$\cite{si,Fang2008d}.
However, for iron-chalcogenides, different magnetic orders
 have been observed\cite{Bao2008,Lis2008a}. For example,
the 11 iron-chalcogenides, $FeTe_{1-x} Se_x$\cite{Hsu2008}, which
can achieve the superconducting
transition temperature around 40K under
pressure\cite{Mizuguchi2008}, can display both commensurate and
incommensurate magnetic state.  The ordered magnetic moment of the parent
compound $FeTe$ is about $2.0\mu_b$, significantly larger than
 iron-pnictides as well.  Recently, it has been shown that the magnetic
  state of $FeTe$, which is a bi-collinear antiferromagnetic (BAFM) can be described by a strongly
   frustrated magnetic model including the nearest neighbor (NN) $J_1$,  the next nearest neighbor(NNN) $J_2$ ,
    and the third nearest neighbor (TNN) $J_3$, i.e. the  $J_1-J_2-J_3$ model\cite{Ma2009a,Fang2009b}.
    It has been determined that $J_1$ is ferromagnetic (FM) while $J_2$ and $J_3$ are
    antiferromagnetic  (AFM)\cite{Lip2011}. It is also interesting to note that  in the BAFM state, the $J_1$
  effectively becomes $J_{1a}$ and $J_{1b}$  due to symmetry
  breaking, which also help to stabilize the BAFM phase.
    The values of the magnetic exchange coupling parameters suggest that  FeTe is close to a boundary between an incommensurate magnetic phase
     and the BAFM phase so that the later can lose its stability by an  introduction of a few percentages of additional Fe atoms\cite{Bao2008}.

However, it is still  hotly debated about whether the effective
models with local magnetic exchange couplings are the right models
to describe the magnetism in iron-based superconductors. For
iron-pnictides, there is less debate since the model is simple and
has gained support from  different families of iron-pnictides.
However,  for $FeTe$, the model lacks of independent verification.
Since  the local physics of the 122 iron-chalogenides  should be
similar to
 the one of  $FeTe$,  it is naturally expected that    both families of iron-chalogenides should be described by similar models.

In this paper, we show that  the magnetism in both families of
iron-chacogenides  can be unifiedly understood within the previous model
obtained for $FeTe$.  The 245 ($K_2Fe_4Se_5$) pattern ($\sqrt{5}\times \sqrt{5}$) of the vacancy
orders  in the 122 system are naturally obtained by maximizing
the magnetic energy saving and reducing the magnetic frustration in
the model.  The magnetic order in the 122 systems becomes robust and
the ordered moment is significantly enhanced. The magnetic
transition temperature around 500k  of $K_2Fe_4Se_5$  quantitatively
agrees with the  pure magnetic energy gain, comparing to the magnetic transition temperature around
70K in $FeTe$ where the magnetic frustration is strong. Therefore,
the vacancy ordering may be originated electronically.  We explore
other  possible vacancy   patterns and magnetic  orders associated
with them.  Moreover, we calculate spin wave excitations and
identify their novel features to quantitatively test the proposed
theory.

Before we start to consider the effect of vacancy, let's review  the
$J_1-J_2-J_3$ model,  which was used to describe the magnetism in
$FeTe$\cite{Ma2009a,Fang2009b}.  The Hamiltonian is given  by
 \begin{eqnarray}
 H=\sum_{\alpha}J_{\alpha}\sum_{<ij>_{\alpha}}S_i \cdot S_j
 \end{eqnarray}
 where $\alpha=1, 2, 3$. The BAFM state in FeTe, which breaks the rotation symmetry,
  is stabilized with a coupling to  the small lattice distortion. In the BAFM state, this broken symmetry can generate the anisotropy among $J_1$ and $J_2$ couplings.
For $FeTe$, by fitting the spin wave spectrum, we have obtained the
values of  the magnetic exchange couplings with $J_1S\sim -34mev$,
$J_2 S\sim 22mev$ and $J_3S\sim7mev$\cite{Lip2011}, where $S$ is the
spin of each site. The anisotropy is shown mainly in $J_1$, with $\frac{J_{1a}-J_{1b}}{2}\sim -16 mev$\cite{Lip2011}. The most important feature revealed from the fitting is that the NN
coupling $J_1$ is FM  while the other two, $J_2$ and
$J_3$, are AFM. The sum of these three coupling
parameters is close to zero, which indicates the model is extremely
magnetically frustrated.

\begin{figure}[tb]
\includegraphics[angle=90,width=6cm]{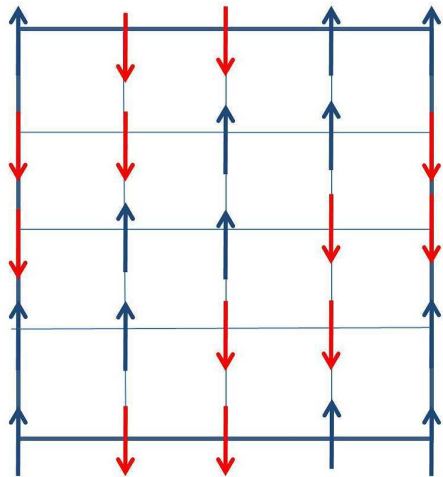}
\caption{The bi-collinear antiferromagnetic phase observed in the 11 (FeTe/Se) systems. \label{bafm}}
\end{figure}

Let's first ignore the  anisotropy  of the magnetic
exchange coupling $J_1$ caused by symmetry breaking in the BAFM
state. We focus on the pure magnetic model in the tetragonal square lattice. In the
BAFM state as shown in fig.\ref{bafm}, both $J_1$ and $J_2$ are
frustrated couplings and the magnetic energy  is only saved by
$J_3$. The saved magnetic energy per site is given by $E_{BAFM}=
-2J_3S^2$. Now, we ask a simple question, what are the possible
stable magnetic orders  if we can remove spins and form vacancy
patterns in  the above model within the similar parameter regions? (note, in the following we
will take $S=1$ for convenience.)

\begin{figure}[tb]
\includegraphics[ width=9cm]{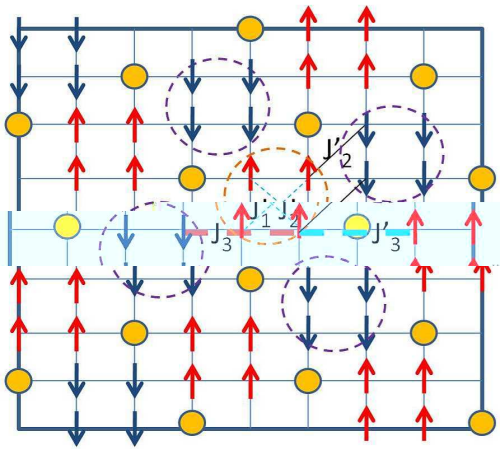}
\caption{The sketch of the 245 vacancy ordering and its magnetic structure. The general magnetic couplings are also indicated.    \label{245}}
\end{figure}

In order to create a vacancy pattern that  minimizes the magnetic exchange energy,  the pattern must sufficiently suppress magnetic frustration and  maximize energy saving from each magnetic exchange coupling simultaneously. It is straightforward to show  that such a pattern is exactly the 245 pattern ( named according to $K_2Fe_4As_5$) as shown in fig.\ref{245}, which has been identified by neutron scattering experiments\cite{Bao2011a}. In the 245 vacancy orders, each site has three NN and NNN couplings to its neighbors. Moreover  one spin in each connected five spins (5-1 pattern) along Fe-Fe direction is removed. The odd number of links for both NN and NN couplings  reduce the magnetic frustration completely. The 5-1 pattern  saves energy from $J_3$ as well. Therefore, the 245 state maximizes the energy saving from  all magnetic exchange couplings. The magnetic energy saving per spin $E_{245}=(-J_1+J_2+2J_3)/2$(we set $J_3=0$ between two NNN sites if there is vacancy between them). If we ignore the lattice distortion and use the exchange parameters derived from $FeTe$, $ E_{245}\sim 70mev$, comparing to $E_{BAFM}\sim 14mev$.  Considering the fact that the ordered magnetic moment is also around 1.5 times larger in $K_2Fe_4As_5$  than in $FeTe$,  the ratio of the magnetic energy savings between two states is given by $1.5\times E_{245}/E_{BAFM}\sim 7.5$.  This ratio is in a quantitative agreement with the ratio of their magnetic transition temperatures $500k/70k\sim 7.1$. (Note: in general, the magnetic transition temperature in a quasi two dimensional material is  given by
$T_N\propto \frac{E}{ln\frac{E}{J_c}}$, where $J_c$ is the coupling strength between layers and $E$ is the in-plane  magnetic energy saving.)  The vacancy pattern is hence a result of the combination of energy saving from three magnetic couplings including the large contribution from the NN ferromagnetic exchange coupling.
If we consider the symmetry breaking of the state which can modify the magnetic coupling strength as shown in fig.\ref{245}, the total energy saving in general can be written as  $E'_{245}=(-2J_1-J'_1+2J'_2-J_2+2J_3-J'_3)/2$ where the parameters are defined in fig.\ref{245}.

\begin{figure}[tb]
\includegraphics[width=9cm]{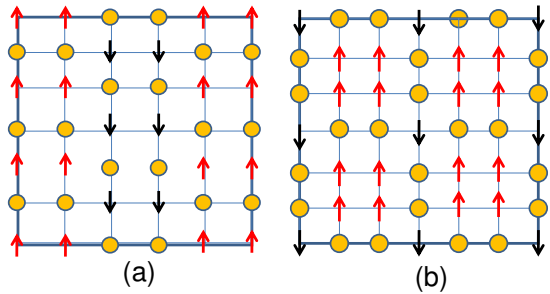}
\caption{(a) The armchair dimer crystal pattern. (b)The square dimer crystal pattern. \label{armchair} }
\end{figure}

\begin{figure}[tb]
\includegraphics[width=8cm]{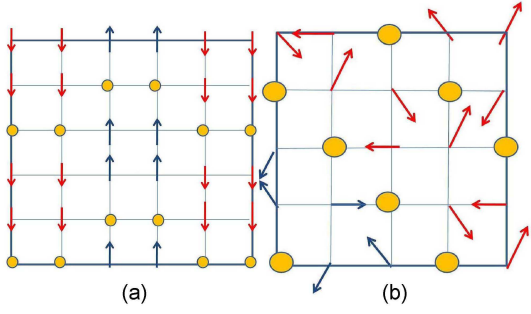}
\caption{ The two different possible patterns for $Fe_{1.33}Se_2$:
(a) The dimer vacancy ordering similar to CAFM magnetic phase. (b)
The diagonal stripe pattern  where the magnetic ordering is
incommensurate. \label{stripe}}
\end{figure}

It is also interesting to explore other possible vacancy patterns and their associated magnetic orders. For such an frustrated $J_1-J_2-J_3$ magnetic model, there is no other pattern, except the 245 ordering, which can get rid of the magnetic frustration completely. With the   values of the exchange couplings given above, we list some possible patterns (we again set $J_3=0$ between two NNN sites if there is vacancy between them): (1) Armchair dimer  crystal patterns (the 212 pattern if we consider $K_2FeSe_2$) shown in fig.\ref{armchair}(a).  In this pattern, half of spins are replaced by vacancies. The magnetic energy saving per site is given by $E_{ADC}=J_2-J_1/2$. The pattern saves both energy from the NN FM and NNN AFM couplings;
(2) Square dimer crystal patterns shown in fig.\ref{armchair}(b).  The energy saving from this structure stems from the NN FM $J_1$ and the NNN AFM $J_2$, $E_{SDC}=\frac{-4J_1+2J_2}{5}$;
\begin{figure}[tb]
\includegraphics[width=4cm]{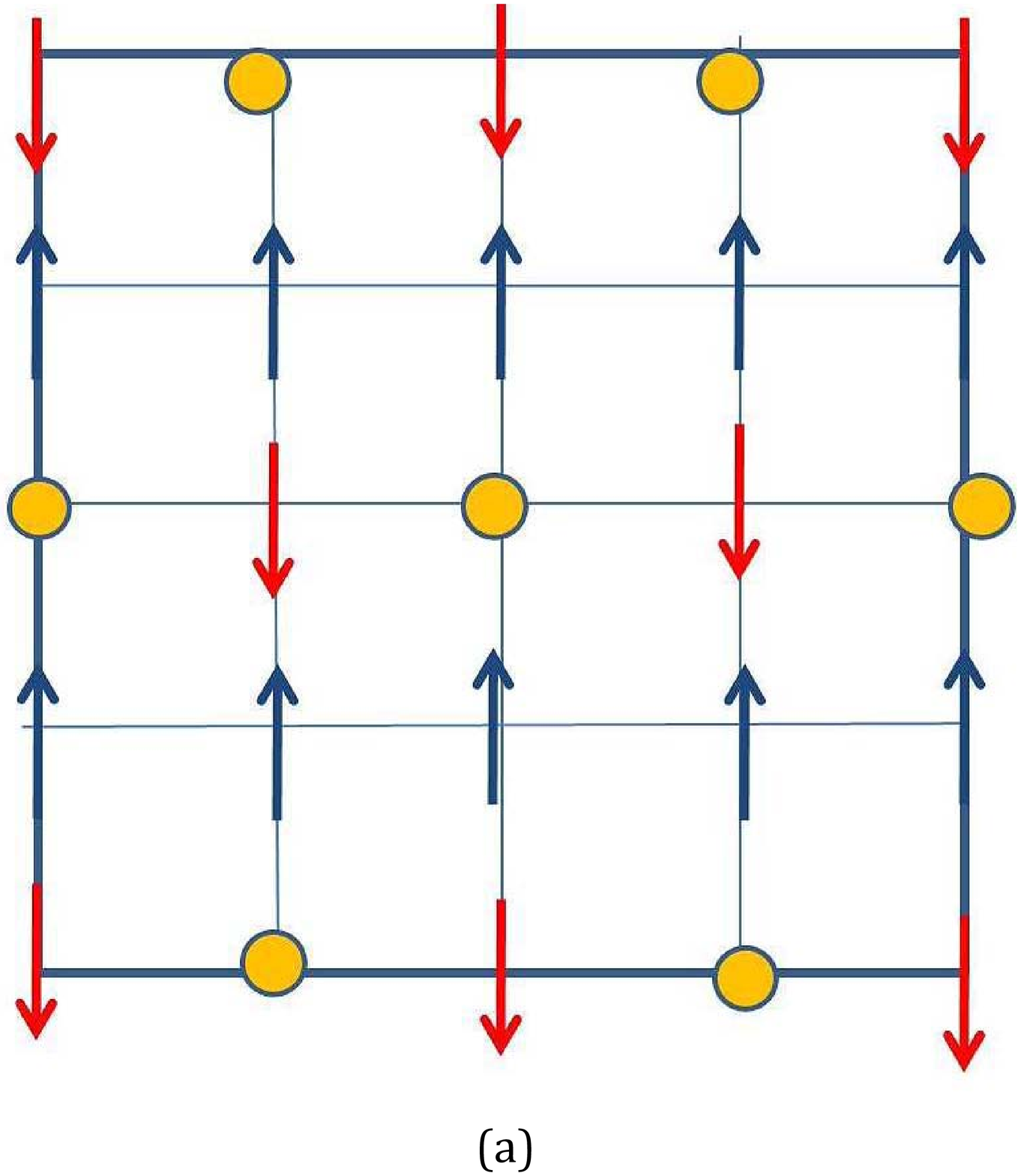}
\includegraphics[width=4cm]{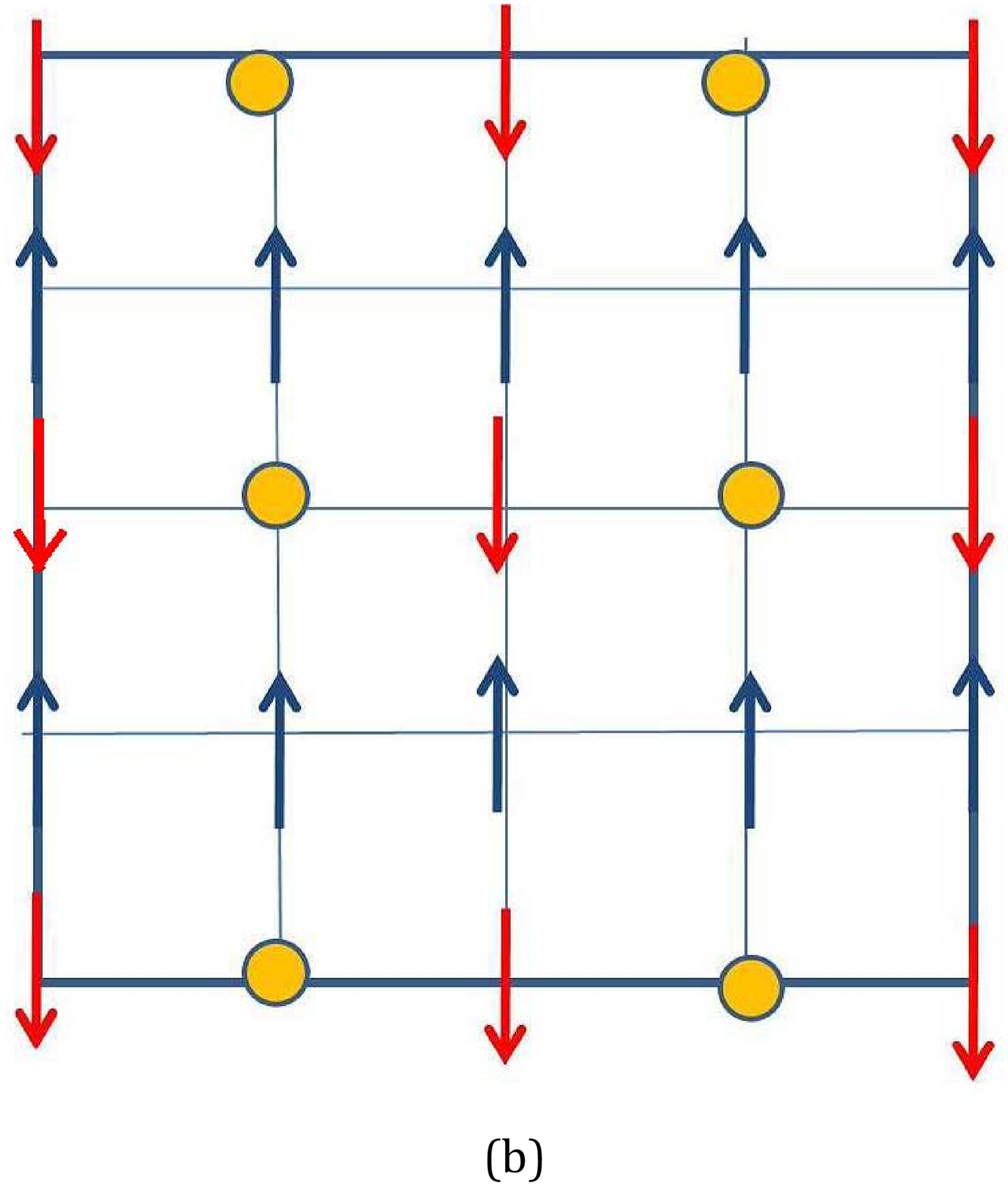}
\caption{  The two different possible patterns (234) for  $Fe_{1.5} Se_2$.\label{234}}
\end{figure}
(3) The 234 vacancy patterns (named as $K_2Fe_3Se_4$) as shown in fig.\ref{234}.  There are two different  patterns.  The first pattern saves energy from $J_1$ and $J_2$ but  pays energy cost  for $J_3$.  The energy saving per site is given by $E_{234a}= -J_1/3+2J_2-2J_3$. This magnetic structure is similar to the $(0,\pi)$ collinear AFM (CAFM) observed in the parental compounds of iron-pnictides.  In the second pattern,  the magnetic state is also similar to CAFM. The energy saving is given by $E_{234b}=\frac{4}{3}(J_2-J_3/2)$.  In general,  the 234 phase where the iron concentration is close to 1.5 is most likely supporting a CAFM magnetic phase, similar to iron pnictides. In both patterns, there are ferromagnetic moments for each layer. Between layers, they are antiferromagnetically coupled. The 234 vacancy patterns have   been observed experimentally\cite{li}. However, the magnetic order has not been identified. (4)  The 446 pattern ($K_4Fe_4Se_6$) as shown in fig.\ref{stripe}.  There are also two possible patterns. The first pattern  is a dimer vacancy ordering.  The magnetic pattern is also  close to CAFM phase with an energy saving given by $E_{234a}=\frac{J_2-J_3-3J_1}{4}$. The second pattern can be viewed as a diagonal stripe similar to undoped cuprates\cite{Emery1999}.  For such a stripe pattern, there is no $J_3$ coupling.  The magnetic structure can be predicted to be incommensurate along the stripe direction and AFM  along  the direction perpendicular to the stripe.  The total energy saving per site is given by
$E_{446}=\frac{3}{2}J_2+\frac{J_1^2}{8J_2}$.  The spin angle between two NN sites along the stripe direction is given by $cos\theta=-\frac{J_1}{4J_2}$.

{\it Properties of the 245 state in the large S limit:} The 245 state is clearly the most stable magnetic state in our model.  In the following, we focus on this state. First, we analyze the spin excitation in the large S limit.  We consider a general model as shown in  Fig.\ref{245}. To start a spin wave calculation, we denote the spin sites as follows. A generic position of the spin is given by $ \mathbf{r}=m\mathbf{l_1}+n\mathbf{l_2}+\mathbf{d_i}$, where $m,n$ are integers and $\mathbf{l_1}=(2\mathbf{x}-\mathbf{y})/\sqrt{5},
\mathbf{l_2}=(\mathbf{x}+2\mathbf{y})/\sqrt{5},$
$ \mathbf{d_1}=0,\;\mathbf{d_2}=\mathbf{x},\;\mathbf{d_3}=\mathbf{x}+\mathbf{y},\;\mathbf{d_4}=\mathbf{y},$ where $\mathbf{x, y}$ are unit vectors in the original tetragonal lattice.
We take the Holstein-Primakoff transform for the given ground state:\\
For $m+n=$even:\bea S_+(\mathbf{r})&=&\sqrt{2S}a_i(\mathbf{R}),\\
\nonumber S_-(\mathbf{r})&=&\sqrt{2S}a^\dag_i(\mathbf{R}),\\
\nonumber S_z(\mathbf{r})&=&S-a^\dag_i(\mathbf{R})a_i(\mathbf{R});\eea
For $m+n=$odd:\bea S_+(\mathbf{r})&=&\sqrt{2S}a^\dag_i(\mathbf{R}),\\
\nonumber S_-(\mathbf{r})&=&\sqrt{2S}a_i(\mathbf{R}),\\
\nonumber S_z(\mathbf{r})&=&-S+a^\dag_i(\mathbf{R})a_i(\mathbf{R}).\eea
The full Hamiltonian of the model can be put in a matrix form. Define $\psi^\dag(k)=(a^\dag_1(k),a^\dag_2(k),a^\dag_3(k),a^\dag_4(k),a_1(-k),a_2(-k),a_3(-k),a_4(-k))$, and we have\bea H=\frac{1}{2}\sum_k\psi^\dag(k)\left(
                                 \begin{array}{cc}
                                   A(k) & B(k) \\
                                   B(k) & A(k) \\
                                 \end{array}
                               \right)
\psi(k).\eea
\begin{widetext}
$A(k)$ and $B(k)$ are four-by-four matrices, defined by:\bea A(k)=S\left(
                                                                                   \begin{array}{cccc}
                                                                                     E_0 & J_1e^{ik_x} & J_2e^{ik_x+ik_y}+J'_3e^{-i2k_x} & J_1e^{ik_y}\\
                                                                                      .& E_0 & J_1e^{ik_y} & J_2^{-ik_x+ik_y}+J'_3e^{-2ik_y} \\
                                                                                      .& . & E_0 & J_1e^{-ik_x}\\
                                                                                      .& . & . & E_0 \\
                                                                                   \end{array}
                                                                                 \right),\eea
\bea B(k)=S\left(
                                                                                   \begin{array}{cccc}
                                                                                      0 & J'_2e^{-ik_x+ik_y}+J_3e^{-2ik_y} & J'_1e^{-ik_y} & J'_2e^{-ik_x-ik_y}+J_3e^{2ik_x}\\
                                                                                      .& 0 & J'_2e^{-ik_x-ik_y}+J_3e^{2ik_x} & J'_1e^{ik_x} \\
                                                                                      .& . & 0 & J'_2e^{ik_x-ik_y}+J_3e^{2ik_y}\\
                                                                                      .& . & . & 0 \\
                                                                                   \end{array}
                                                                                 \right),\eea
                                                                                 \end{widetext}
where $E_0=-2J_1-J_2+J'_1+2J'_2+2J_3-J'_3$. The lower triangle elements are suppressed because both matrices are hermitian.

By diagonalizing this Hamiltonian,
 we have\bea H=\sum_{i=1,2,3,4;k}(\gamma^\dag_i(k)\gamma_i(k)+1/2)\omega_i(k),\eea and\bea a_i(k)=\sum_j U_{ij}(k)\gamma_j(k)+V_{ij}(k)\gamma^\dag_j(-k).\eea where $U$ and $V$ are $4\times4$ matrices to diagonalize the Hamiltonian.  The spin reduction due to quantum fluctuations on each site is given by:\bea\delta{m}=\frac{1}{4\pi^2}\sum_{j}\int_0^{2\pi}dk_x\int_0^{2\pi}dk_y|V_{1j}(k_x,k_y)|^2.\label{eq:moment_reduction}\eea It is also interesting to calculate the dynamic factor $S(\omega)=-\Im[\chi(\omega)]$, where the local susceptibility $\chi(\omega)$ is defined as \bea
 & & \chi(i\omega_n)=\sum_q\int{d\tau}e^{i\omega_n\tau}[\langle S^+(q,\tau)S^-(-q,0)\rangle \nonumber \\
& & +\langle S^-(q,\tau)S^+(-q,0)\rangle]/N.\eea After taking Fourier transform, we obtain
\bea
& & \chi(\omega)=\sum_k\sum_{i,j}(|V_{i,j}(k)|^2+|U_{i,j}(k)|^2)\nonumber \\
& & (\frac{1}{\omega-E_j(k)+i0}+\frac{1}{\omega+E_j(k)+i0})/N.\eea

Due to the enlarged unit cell, the spinwave BZ (SBZ) is smaller than the original unfolded BZ for one Fe per unit cell. The SBZ is cornered by the following four (equivalent) points $(-3\pi/5,-\pi/5)$, $(\pi/5,-3\pi/5)$, $(3\pi/5,\pi/5)$, and $(-\pi/5,3\pi/5)$, and these points are called $M_s$ points, different from the real $M$ point in lattice BZ. It is convenient to define $X_s=(2\pi/5,-\pi/5)$ and $Y_s=(-\pi/5,-2\pi/5)$ for future discussion.
\begin{figure}[tb]
\includegraphics[width=8cm]{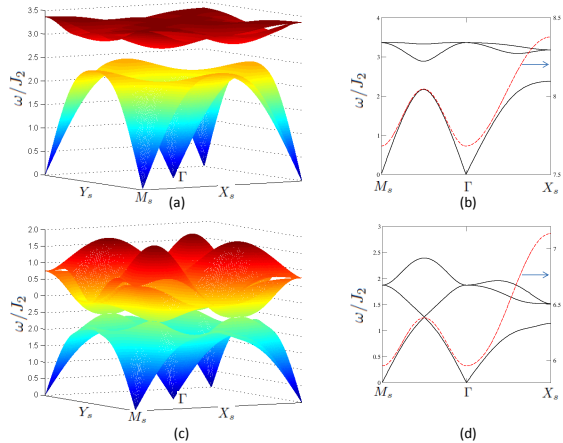}
\caption{The spinwave spectra of the lowest three branches (the other one is too high to be drawn in the same plot) using the parameters $(J_1,J'_1,J_2,J'_2,J_3,J'_3)=(-30,-10,20,20,9,0)$meV in (a) and $(J_1,J'_1,J_2,J'_2,J_3,J'_3)=(-30,-30,20,20,9,0)$meV in (b). The corresponding band structures are given in (b) and (d), in which the read dashed lines represent the highest branch and black solid lines the lowest three branches. $M_s$, $X_s$ and $Y_s$ are high symmetry points in the spinwave BZ that are defined in text.\label{fig:spectra}}
\end{figure}
The spinwave has four branches. Depending on the parameters, there can be finite gaps between the first and the second branches and between the third and the fourth ones . The middle two branches are degenerate at $\Gamma$ and $M_s$ points and are in general close to each other. At high symmetry points, the spinwave energy has analytic expressions. We take $S=1$ for convenience. At $M_s$ and $\Gamma$: $ E_1(\Gamma)=0,
 E_2(\Gamma)=E_3(\Gamma)=2\sqrt{(J_e-J_1-J_2 -J'_3)(J'_1-J_1-J_2+J_e-J'_3)}$, and $
 E_4(\Gamma)=2\sqrt{2}\sqrt{(J'_1-2J_1)(J_e-J_1)}$ where $J_{e}=J'_2+J_3$ are the effective antiferromagnetic coupling strength between two 4-spin plaquettes.  At $X_s$,
  $E_1(X_s)=2(J_1+\sqrt{(J_e-J'_3-J_1)(J'_1+J_e-J'_3-J_1)}),
 E_2(X_s)=2\sqrt{J_1^2+(J_2-2J'_2)(J_2-2J_3)+\omega}$, where $\omega= J'_1(J_e-J_2)-J_1(J'_1+2J_e-2J_2)$, and
  $E_4(X_s)=2(-J_1+\sqrt{(J_e-J'_3-J_1)(J'_1+J_e-J'_3-J_1)})$, which also defines the spinwave  band width. The first spinwave branch effectively describes the spin fluctuations of 4-spin plaquettes.

  We calculate spinwave spectra and related features for three different parameter sets. We first adopt the parameters: $J_1=(J_{1a}+J_{1b})/2=-30$meV, $J_2=20$meV, $J_3=9$meV. Due to lattice distortion that draws the four spins in each new unit cell closer, the primed parameters can be reduced from unprimed ones by the presence of iron vacancies: $J'_1=-10$meV, $J'_2=20$meV and $J'_3=0$meV. For these parameters the spinwave spectra have a finite gap $\Delta_{12}=10.2$meV between the first and the second branches, and a larger gap $\Delta_{34}=86.2$meV between the third and fourth branches. The full bandwidth is $W=167.5$meV. The first branch starts from zero energy at $\Gamma$ and $M_s$ and reaches maximum at $X_s$; the second and third branches start from minimum at $(3\pi/10,\pi/10)$ and its three $C_4$ equivalents, and reach maximum at $(0.45\pi,0.12\pi)$. The fourth branch starts from $E_4(\Gamma)$ at $\Gamma$ and $M_s$ to $E_4(X_s)$ at $X_s$. The ordered moment correction from spinwave is $\delta{m}=0.095$. See Fig.\ref{fig:spectra}(a,b) for details.

In choosing the second parameter set, we consider a purely magnetic model without lattice distortion. Therefore we take $J'_1=J_1$, $J'_2=J_2$. However one still has $J_3\gg J'_3\sim0$ because of the vacancy along the exchange path for $J'_3$. For this parameter set, there is no gap between the first and second branches, but the gap between third and fourth branch remains finite. The full bandwidth in this case is $W=142.7$meV. The first three branches start from zero energy at $\Gamma$ and $M_s$, and reach maximum at $(3\pi/10,\pi/10)$ and its three $C_4$ equivalents. The fourth branch's minimum and maximum appear at the same points as in the first parameter set. The ordered moment correction from spinwave is $\delta{m}=0.197$. See Fig.\ref{fig:spectra}(c,d) for details.

In FeTe, we found that the third neighbor exchange $J_3$ is needed to explain both low and high energy regions of dynamic spin susceptibility observed in neutron experiment, while a parameter fit without $J_3$ significantly underestimates the spin wave bandwidth. $J_3$ is also important in the current system to stabilize the 245 state. For comparison, we adopt the parameters we used to fit FeTe system without $J_3$ (see supplementary materials in Ref.\cite{Lip2011}): $(J_1,J'_1,J_2,J'_2,J_3,J'_3)=(-12,-4,9,9,0,0)$meV. The spinwave correction to moment in this case is $\delta{m}=0.43$, indicating much stronger spin fluctuation.

\begin{figure}[!htb]
\includegraphics[width=8cm]{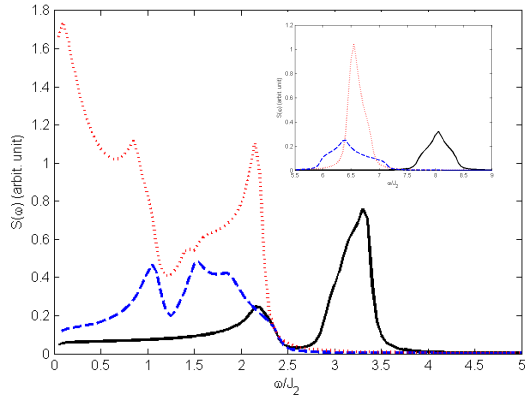}
\caption{The dynamic factor $S(\omega)$ contributed by spinwave. Black solid line corresponds to $(J_1,J'_1,J_2,J'_2,J_3,J'_3)=(-30,-10,20,20,9,0)$meV; blue dashed line corresponds to $(J_1,J'_1,J_2,J'_2,J_3,J'_3)=(-30,-30,20,20,9,0)$meV; and red dotted line $(J_1,J'_1,J_2,J'_2,J_3,J'_3)=(-12,-4,9,9,0,0)$meV. The main panel only includes contribution from lowest three branches as the highest branch is too high to be plotted in a same energy range. The inset shows the contribution by the fourth and highest branch.\label{fig:dynamic}}
\end{figure}

The dynamic factor is shown in Fig.\ref{fig:dynamic}. Note that around zero energy the dynamic factor should drop to zero in a 3D system, but converge to a finite value in our 2D system. If we include a small $J_z$ in the beginning, we can make the curve drop to zero at zero energy, however the dynamic factor at $\omega\gg J_z$ is unaffected. See Fig.\ref{fig:dynamic} for the dynamic factor with all three parameter sets given above.

\begin{figure}[tb]
\includegraphics[width=6cm]{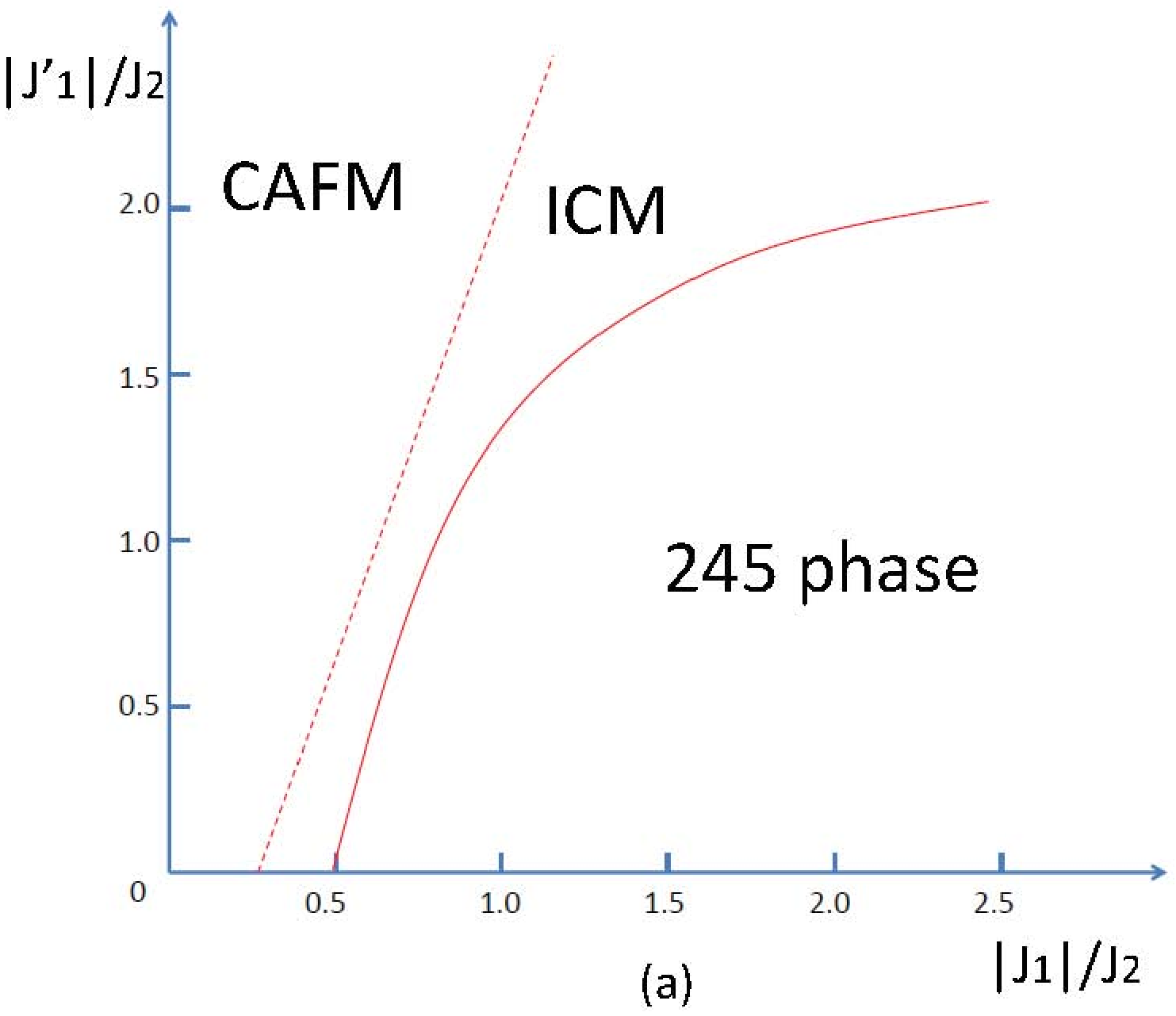}
\includegraphics[width=6cm]{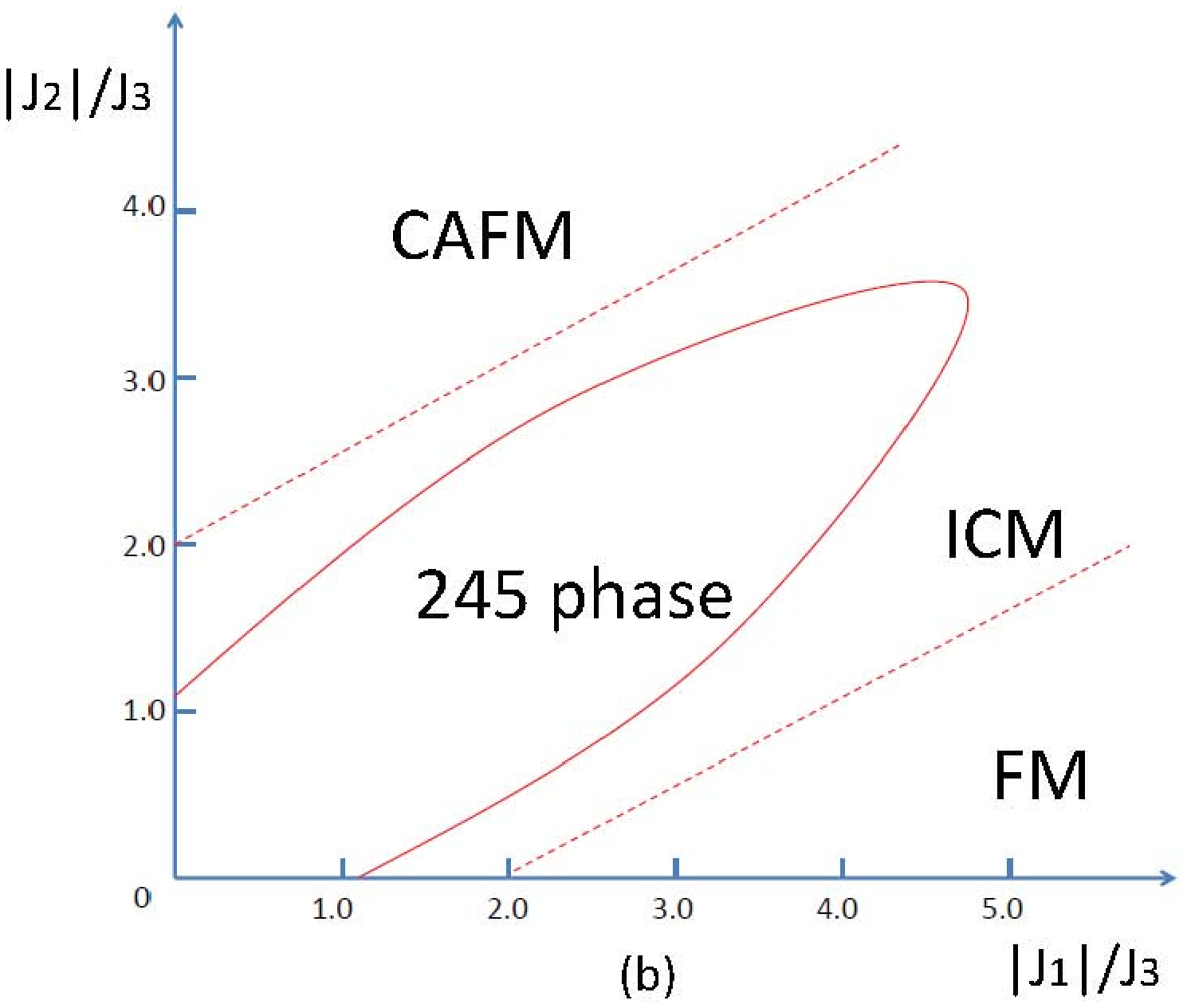}
\caption{The phase diagrams against $J_1$ and $J'_1$ (left) and $J_1$ and $J_2$ (right). Solid lines mark second order transitions calculated from spinwave and dotted lines mark possible first order transitions. \label{fig:phase}}
\end{figure}

It is also interesting to check what is the closest magnetic state if the exchange coupling parameters can be varied in the 245 state. In the reasonable parameter space, the following states can  be considered: 1. The ferromagnetic state if $J_1$ and $J'_1$ are large. 2. the CAFM state when  $J_2$ is large. 3. The incommensurate phase. The ground state energy for the first two are easy to calculate: $E_{FM}=4J_1+2J'_1+2J_2+4J'_2+4J_3+2J'_3$ and $E_{AFM2}=-2J_2-4J'_2+4J_3+2J'_3$. Comparing the energies, we have if $|J'_1|>2(J'_2+J_3)$, the FM phase wins over the 245 magnetic phase; and if $2|J_1|-|J'_1|<2J_2-4J_3$, CAFM state has lower energy than the 245 magnetic state and thus becomes preferable. Most importantly, an ICM state is the closest competitor to the 245 magnetic phase. An analysis of the spinwave Hamiltonian around $\Gamma$ point gives us the phase boundary between the 245 magnetic state  and ICM as drawn in fig.\ref{fig:phase}.  The boundary is set by the equation: $  J_1(J'_1+2(J_e-J'_3))+J'_1(J_2-J_e+J'_3)+2(J_2(J_e-J'_3)+2J_3(-J'_2+J'_3)=0$.  We draw  fig.\ref{fig:phase} (left) for a phase diagram against $J_1$ and $J'_1$, fixing $J_2=J'_2$ and $J'_3=9J_2/20$, and  fig.\ref{fig:phase} (right) for a phase diagram against $J_1$ and $J_2$, fixing $J'_1=J_1$, $J'_2=J_2$ and $J'_3=0$.

{\it Discussion:} We have shown that the 245 vacancy pattern can be naturally obtained  through saving magnetic energy  in a strongly frustrated magnetic model $J_1-J_2-J_3$ model.  A few remarks regarding our theory need to be addressed. First,
in our above analysis, we only consider the magnetic energy and ignore other possible energy sources. In real materials, it is possible that other energy sources may make a significant contribution as well. For example,
 creating a dimer of two vacancies may cost higher or lower energy than being two individual vacancies.
 Therefore, comparing  a dimer vacancy pattern  to a non-dimer pattern requires further careful consideration.
 The LDA calculation may help to address it\cite{Yan2011}. Second, we expect a lattice distortion to naturally take place
 above or equal to the magnetic transition temperature, similar to what has been observed in iron-pnictides\cite{Cruz2008}.
 Third, some LDA calculations have predicted the values of magnetic exchange couplings for $K_2Fe_4As_5$\cite{Cao2010}. However, their values are not consistent with our picture. In their results, the NN exchange coupling $J'_1$ is antiferromagnetic and there is no $J_3$.  We expect that both $J_1$ and $J'_1$ should be FM and $J_3$ should be significant.
 Third, it is very important to notice that the NN exchange coupling is strongly ferromagnetic,
 which contributes the largest energy in the 245 phase.  It leads to a high magnetic transition temperature observed in the 245 phase.
 It is interesting to note that the 245 pattern has also
  been obtained in one band $t-J-V$ model\cite{Kivelson1990} which is  driven by magnetic energy saving in proper parameter regions.   Fourth,
  the ferromagnetic coupling does not contribute to any spin singlet pairing in superconducting (SC) state.  Therefore, although
   the magnetic transition temperature is high in these new materials,  we {\it do not} expect the SC transition temperature
   can be scaled as the magnetic transition temperature here because  the SC transition temperature is   determined by
    the strength of the AFM couplings. Fifth, the SC  pairing is mainly determined by $J_2$\cite{seo2008}. We expect the $J_2$ value is similar to the one in iron-pnictieds\cite{Zhaoj2009,Zhaojun2008} since the SC transition temperatures for both materials are similar. Sixth, the existence of $J_3$ distinguishes iron-chacogenides from iron-pnictides and generate
     many different magnetic states. The effect of $J_3$ in superconducting state will be addressed in a different paper.
      In fact, $J_3$ can enhance the s-wave pairing significantly when electron pockets  dominate over hole pockets. Finally,
       the 245 pattern is so strong in magnetism. We do not think this state can be coexisted with superconducting phase.

In summary, the $J_1-J_2-J_3$ model provides a natural and unified understanding for  magnetism and vacancy ordering in iron chacogenides.  The 245 vacancy   pattern reduces the magnetic frustration and achieves the maximum magnetic energy saving. The reduction of frustration increases the ordered magnetic moment as well as strongly enhancing the magnetic transition temperature. (Note, after finishing our paper, we find   similar spinwave excitations for the 245 pattern are  calculated and discussed in\cite{Youy2011} although the parameters in their model and the physical motivation are very different from us.)

{\it Acknowledgement:} JP thanks  S. A. Kivelson, Hong Ding, Donglei
Feng and X.H. Chen for useful discussion.

\bibliographystyle{prsty}
\bibliography{myself3,hightc,iron5,paper,fese2}

\end{document}